\documentclass[aps,prl,twocolumn,groupedaddress]{revtex4-1}

\usepackage{amsmath,amsthm,amssymb}
\usepackage{mathtools}
\usepackage{enumitem} 
\usepackage{graphicx}

\begin{document}

\title{Not all disadvantages are equal: Racial/ethnic minority students have largest disadvantage of all demographic groups in both STEM and non-STEM GPA}

\author{Kyle~M.~Whitcomb}
\affiliation{Department of Physics and Astronomy, University of Pittsburgh, Pittsburgh, PA, 15260}
\author{Chandralekha~Singh}
\affiliation{Department of Physics and Astronomy, University of Pittsburgh, Pittsburgh, PA, 15260}

\date{\today}

\begin{abstract}
	An analysis of institutional data to understand the outcome of the many obstacles faced by students from historically disadvantaged backgrounds	is important in order to work towards promoting equity and inclusion for all students.
	We use 10 years of institutional data at a large public research university to investigate the grades earned (both overall and in STEM courses only) by students categorized on four demographic characteristics: gender, race/ethnicity, low-income status, and first-generation college student status.
	We find that on average across all years of study and for all clusters of majors, underrepresented minority students experience a larger penalty to their mean overall and STEM GPA than even the most disadvantaged non-URM students.
	Moreover, the underrepresented minority students with additional disadvantages due to socioeconomic status or parental education level were even further penalized in their average GPA.
	Furthermore, we also find that while women in all demographic groups had a higher average overall GPA, these gender differences are almost completely non-existent in STEM GPA except among the most privileged students.
	These findings suggest that there is need to provide support to bridge the gaps that emanate from historical disadvantages to certain groups.
\end{abstract}

\maketitle

\section{Introduction and Theoretical Framework}

The importance of evidence-based approaches to improving student learning and ensuring that all students have the opportunity to excel regardless of their background is becoming increasingly recognized by Science, Technology, Engineering, and Mathematics (STEM) departments across the US~\cite{johnson2012, johnson2017, maltese2011, borrego2008, borrego2011, borrego2014, henderson2008, dancy2010, henderson2012}. 
With advances in digital technology in the past few decades, institutions have been keeping increasingly large digital databases of student records.
We have now reached the point where there is sufficient data available for robust statistical analyses using data analytics that can provide valuable information useful for transforming learning for all students~\cite{baker2014, papamitsiou2014}. 
This has lead to many recent studies utilizing many years of institutional data to perform analyses that were previously limited by statistical power~\cite{lord2009, lord2015, ohland2016, matz2017, witherspoon2019}.
Therefore, here we focus on harnessing institutional data to investigate the obstacles faced by students with various disadvantages who must overcome obstacles in their pursuit of higher education.

The theoretical framework for this study has two main foundations: critical theory and intersectionality.
Critical theories of race, gender, etc. identify historical sources of inequity within society, that is, societal norms that perpetuate obstacles to the success of certain groups of disadvantaged people~\cite{crenshaw1995, kellner2003, yosso2005, gutierrez2009, taylor2009, tolbert2018, schenkel2020}. 
Critical theory tells us that the dominant group in a society perpetuates these norms, which are born out of their interests, and pushes back against support systems that seek to subvert these norms~\cite{crenshaw1995, kellner2003, yosso2005}. 
These highly problematic societal norms are founded in the historical oppression of various groups of people, and manifest today in many ways including economic disadvantages, stereotypes about who can succeed in certain career paths, and racist and/or sexist barriers to opportunity, including educational advancement.
While these norms are, by definition, specific to a particular culture or even country, they are nonetheless pervasive and oppressive and demand attention to rectify these historical wrongs.

Much important work has been done on building critical race and/or gender theories of STEM education~\cite{johnson2012, johnson2017, solorzano2000, lewis2009, bang2010, estrada2018, ong2018, tolbert2018, green2019, mutegi2019, sheth2019, schenkel2020}. 
In one study, Bancroft (2018) lays out a ``critical capital theory,'' using varying forms of capital (economic, social, and cultural) to examine persistence through graduation in STEM doctoral programs and to contextualize the mechanisms behind racial inequities in STEM education~\cite{bancroft2018}.
The idea that race, gender, or another demographic characteristic alone cannot fully explain the intricacies of the obstacles that students face is rooted in the framework of intersectionality~\cite{crenshaw1990, cho2013, mitchell2014, charleston2014, morton2018}. 
In particular, the combination of different aspects of an individual's social identity (e.g., gender, race, first-generation college status, and socioeconomic status) leads to unique levels of disadvantages that cannot be explained by simply adding together the effects of the individual components of identity~\cite{crenshaw1990}. 
For example, according to the framework of intersectionality, in many STEM disciplines where the societal norm expects that students are white men, the experience of a black woman is not a simple sum of the experiences of white women and black men~\cite{charleston2014, morton2018}. 

With an eye toward this intersectional approach to critical theory, we seek to understand the relationship between four different aspects of student identity that can lead to obstacles in STEM education: race/ethnicity, gender, low-income status, and first-generation college student status.
The students disadvantaged by low-income or first-generation status are likely to experience a lack of resources relative to their more privileged peers~\cite{lam2005, dika2016, aruguete2017}. 
Women and underrepresented minority students are susceptible to additional stress and anxiety from stereotype threat (i.e., the fear of confirming stereotypes pertaining to their identity) which is not experienced by their majority group peers~\cite{lewis2009, johnson2012, green2019, mutegi2019, sheth2019, astin1993, cross1993, felder1995, felder1998, bianchini2002, britner2006, bianchini2013, basile2015, cheryan2017, hilts2018}. 
In summary, the different mechanisms by which students belonging to each demographic characteristic can be disadvantaged are as follows.
\begin{itemize}
	\item Race/Ethnicity: Students belonging to underrepresented minority (URM) groups may experience stereotype threat that causes anxiety and robs the students of their cognitive resources, particularly during high-stakes testing.
	\item Gender: There are pervasive societal biases against women succeeding in many STEM disciplines which can result in stereotype threat.
	\item Low-Income Status: Low-Income (LI) students are more likely to need to work to support themselves, reducing their time and energy available to devote to their studies, in addition to anxiety due to the financial burden of attending college. These burdens are in addition to other factors that low-income students may be more likely to face, such as lower quality preparation for college.
	\item First-Generation Status: First-Generation (FG) students may lack the resources of encouragement, advice, and support that are available more readily to students with degree-holding parents. This lack of resources can make FG students more susceptible to the stress of the unknown in college.
\end{itemize}
All of these mechanisms can produce an inequitable learning environment wherein students belonging to any of these groups are forced to work against obstacles that their peers do not have.
The framework of intersectionality asserts that for students that belong to more than one of these groups, complex interactions between these different obstacles can result in compounded disadvantages that are not a simple sum of the individual effects~\cite{crenshaw1990, cho2013, mitchell2014, charleston2014, morton2018}. 

In order to measure the long-term effects of these systemic disadvantages, we will investigate the academic achievement of students belonging to these various demographic groups over the course of their studies at one large public research university using 10 years of institutional data.
By grouping students according to their demographic background, we will be able to investigate how different combinations of obstacles affect student grade point averages.

\section{Research Questions}

Our research questions regarding the intersectional relationships between demographic characteristics and academic achievement are as follows.

\begin{enumerate}[label={\bfseries RQ\arabic*.}, ref={\bfseries RQ\arabic*}, itemsep=1pt]
	\item \label{rq_demo} Are there differences in the overall or STEM grades earned by students belonging to different demographic groups (i.e., underrepresented minority, low-income status, and first-generation college student status)?
	\item \label{rq_gender} Do any patterns observed in RQ1 differ for men and women?
	\item \label{rq_stem} Do grades earned in STEM courses alone exhibit similar demographic patterns as grades earned in all courses?
	\item \label{rq_time} What are the trends over time in the mean GPA of these different demographic groups among different clusters of majors (i.e., computer science, engineering, mathematics, and physical science majors, other STEM majors, and non-STEM majors)?
\end{enumerate}

\section{Methodology} 

\subsection{Sample}

Using the Carnegie classification system, the university at which this study was conducted is a public, high-research doctoral university, with balanced arts and sciences and professional schools, and a large, primarily residential undergraduate population that is full-time and reasonably selective with low transfer-in from other institutions~\cite{carnegie}.

The university provided for analysis the de-identified institutional data records of students with Institutional Review Board approval.
In this study, we examined these records for $N = 24,567$ undergraduate students enrolled in three colleges within the university: the colleges of Arts and Sciences, Computing and Information, and Engineering.
This sample of students includes all of those from ten cohorts who met several selection criteria, namely that the student had first enrolled at the university in a Fall semester from Fall 2005 to Fall 2014, inclusive, and the institutional data on the student was not missing or unspecified for any of the following measures: gender, race/ethnicity, parental education level, and family income.
This sample of students is $50\%$ female and had the following race/ethnicities: 79\% White, 9\% Asian, 7\% Black, 3\% Hispanic, and 2\% other or multiracial.
Further, this sample is $16\%$ first-generation college students and $21\%$ ``low-income'' students (to be defined in the following section).

We acknowledge that gender is not a binary construct, however in self-reporting their gender to the university students were given the options of ``male'' or ``female'' and so those are the two self-reported genders that we are able to analyze.
There were $39$ students who had met all other selection criteria but who had not indicated any gender on the survey, these students were removed from the sample and are not included in the reported sample size or any analyses.

\subsection{Measures}

\subsubsection{Demographic Characteristics}

Four primary measures are the demographic characteristics mentioned in the previous section, namely gender, race/ethnicity, parental education level, and family income.
All of these were converted into binary categories intended to distinguish between the most and least privileged students on each measure.
\begin{itemize}
	\item \textit{Gender}. Gender was reported as a binary category to begin with (either ``male'' or ``female''), therefore no further steps were required.
	\item \textit{First-generation}. Students for whom both parents had a highest completed level of education of high school or lower were grouped together as ``first-generation'' (FG) college students and correspondingly students for whom at least one parent had earned a college degree were labeled non-FG.
	\item \textit{Low-income}. Students whose reported family Adjusted Gross Income (AGI) was at or below 200\% of the federal U.S. poverty line were categorized as ``low-income'' (LI), and those above 200\% of the poverty line as non-LI~\cite{cauthen2007, jiang2015}.
	\item \textit{Underrepresented minority}. All students who identified as any race or ethnicity other than White or Asian were grouped together as ``underrepresented minority'' (URM) students, including multiracial students who selected White and/or Asian in addition to another demographic option. Students who only identified as White and/or Asian students were categorized as non-URM students.
\end{itemize}

\subsubsection{Academic Performance}

Measures of student academic performance were also included in the provided data.
High school GPA was provided by the university on a weighted scale from 0-5 that includes adjustments to the standard 0-4 scale for Advanced Placement and International Baccalaureate courses.
The data also include the grade points earned by students in each course taken at the university.
Grade points are on a 0-4 scale with $\text{A}=4$, $\text{B}=3$, $\text{C}=2$, $\text{D}=1$, $\text{F}=0$, where the suffixes ``$+$'' and ``$-$'' add or subtract, respectively, $0.25$ grade points (e.g. $\text{B}-=2.75$), with the exception of $\text{A}+$ which is reported as the maximum 4 grade points.
The courses were categorized as either STEM or non-STEM courses, with STEM courses being those courses taken from any of the following departments: biological sciences, chemistry, computer science, economics, any engineering department, geology and environmental science, mathematics, neuroscience, physics and astronomy, and statistics.
We note that for the purposes of this paper, ``STEM'' does not include the social sciences other than economics, which has been included due to its mathematics-intensive content.

\subsubsection{Year of Study}

Finally, the year in which the students took each course was calculated from the students' starting term and the term in which the course was taken.
Since the sample only includes students who started in fall semesters, each ``year'' contains courses taken in the fall and subsequent spring semesters, with courses taken over the summer omitted from this analysis.
For example, if a student first enrolled in Fall 2007, then their ``first year'' occurred during Fall 2007 and Spring 2008, their ``second year'' during Fall 2008 and Spring 2009, and so on in that fashion.
If a student is missing both a fall and spring semester during a given year but subsequently returns to the university, the numbering of those post-hiatus years is reduced accordingly.
If instead a student is only missing one semester during a given year, no corrections are made to the year numbering.
In this study we consider up through the students' sixth year of study or the end of their enrollment at the studied institution, whichever comes first.

\subsection{Analysis}

The primary method by which we grouped students in this analysis was by their set of binary demographic categories.
This grouping was performed in two different ways.
First, use of all four binary categories (gender, FG, LI, URM) resulted in sixteen mutually exclusive groups (e.g., ``female, FG+URM'' or ``male, LI'').
Second, use of all categories except gender resulted in eight mutually exclusive categories.

We calculated each student's yearly (i.e., not cumulative) grade point average (GPA) across courses taken in each year of study from the first to sixth years.
In addition, we calculated the student's yearly STEM GPA, that is, the GPA in STEM courses alone.
Then, using the aforementioned grouping schemes, we computed the mean GPA in each demographic group as well as the standard error of the mean separately for each year of study~\cite{freedman2007}.
Further, in the case of grouping by gender, we computed the effect size of the gender differences within each demographic group using Cohen's $d$, which is typically interpreted using minimum cutoff values for ``small'' ($d=0.20$), ``medium'' ($d=0.50$), and ``large'' ($d=0.80$) effect sizes~\cite{cohen1988, neter1996, montgomery2012}.

All analyses were conducted using R~\cite{rcran}, making use of the package \texttt{tidyverse}~\cite{tidyverse} for data manipulation and plotting.

\section{Results}

\subsection{GPA Trends by Demographic Group: ``Dinosaur Plots''}

\begin{figure*}
    \centering
    
	\includegraphics[width=0.95\textwidth]{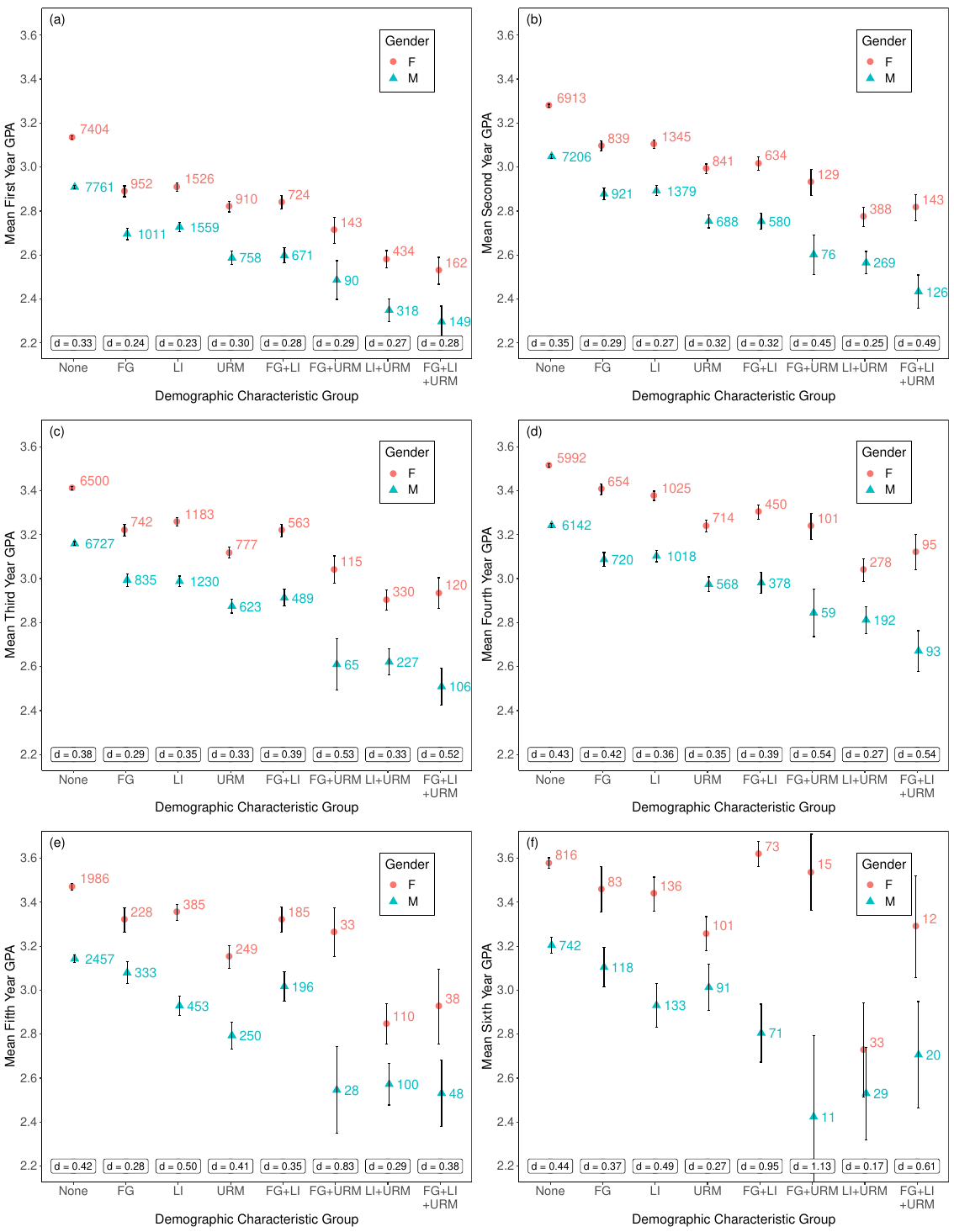}
	
	\caption{\label{figure_gpa_demo_all}
	Average GPA of each demographic group. Students are binned into separate demographic groups based on their status as first-generation (FG), low-income (LI), and/or underrepresented minority (URM) students. The men and women in each demographic group are plotted separately. The mean GPA in all courses taken by students in each demographic group is plotted along with the standard error on the mean, with a separate plot for each of the (a) first, (b) second, (c) third, (d) fourth, (e) fifth, and (f) sixth years. The sample size is reported by each point, and Cohen's $d$~\cite{cohen1988} measuring the effect size of the gender difference in each group is reported.}
\end{figure*}

In order to answer \ref{rq_demo}, we plotted in Fig.~\ref{figure_gpa_demo_all} the mean GPA earned by students in each demographic group, including gender as a grouping characteristic.
We start with overall GPA, rather than STEM GPA alone, in order to provide context for the results in STEM GPA and identify trends that may or may not be present when viewing STEM grades alone.
Groups are ordered from left to right first by the ascending number of selected characteristics and then alphabetically.
Mean GPA is plotted separately (i.e., not cumulatively) for each year of study from the first to sixth year.
Setting aside the gender differences for a moment, we note that the general GPA trends by demographic group in Fig.~\ref{figure_gpa_demo_all} follow a shape resembling the neck, back, and tail of a sauropod, and so accordingly we refer to the plots in Fig.~\ref{figure_gpa_demo_all} as ``dinosaur plots.''
This shape is clearest in the plots for the first through fourth years, as the sample size drops significantly in the fifth year as the majority of students graduate.

Looking more closely at Fig.~\ref{figure_gpa_demo_all}, particularly the first four years, we see that the ``neck'' is consistently comprised of the group of students with the most privileges, namely those students that are non-FG, non-LI, and non-URM.
Following this, the ``back'' is relatively flat across the next four groups, namely students that are FG only, LI only, URM only, or FG and LI.
Notably, the URM group of students typically have the lowest mean GPA within this set of demographic groups.
Finally, the ``tail'' consists of the final three groups, FG+URM, LI+URM, and FG+LI+URM.
The mean GPA in this set of groups tends to decrease from left to right in the plots.
Notably, the four groups that contain URM students are consistently in the lowest four or five mean GPAs.

\subsection{Intersectionality with Gender}

We now turn our attention to the differences between men and women in Fig.~\ref{figure_gpa_demo_all} in order to answer \ref{rq_gender}.
We note in particular that across all demographic groups women's mean GPA is roughly 0.2 grade points higher than men's.
The effect sizes (Cohen's $d$) of this difference range from small to medium~\cite{cohen1988}.
This difference in mean GPA earned is substantial enough to indicate a change in letter grade, given that the grading system at the studied university uses increments of 0.25 grade points for letter grades containing ``$+$'' or ``$-$.''
Further, this trend holds in the fifth year (Fig.~\ref{figure_gpa_demo_all}e) and sixth year (Fig.~\ref{figure_gpa_demo_all}f), with some exceptions in demographic groups with particularly low sample sizes after the fourth year.

\subsection{STEM GPA Trends}

\begin{figure*}
    \centering
    
	\includegraphics[width=0.95\textwidth]{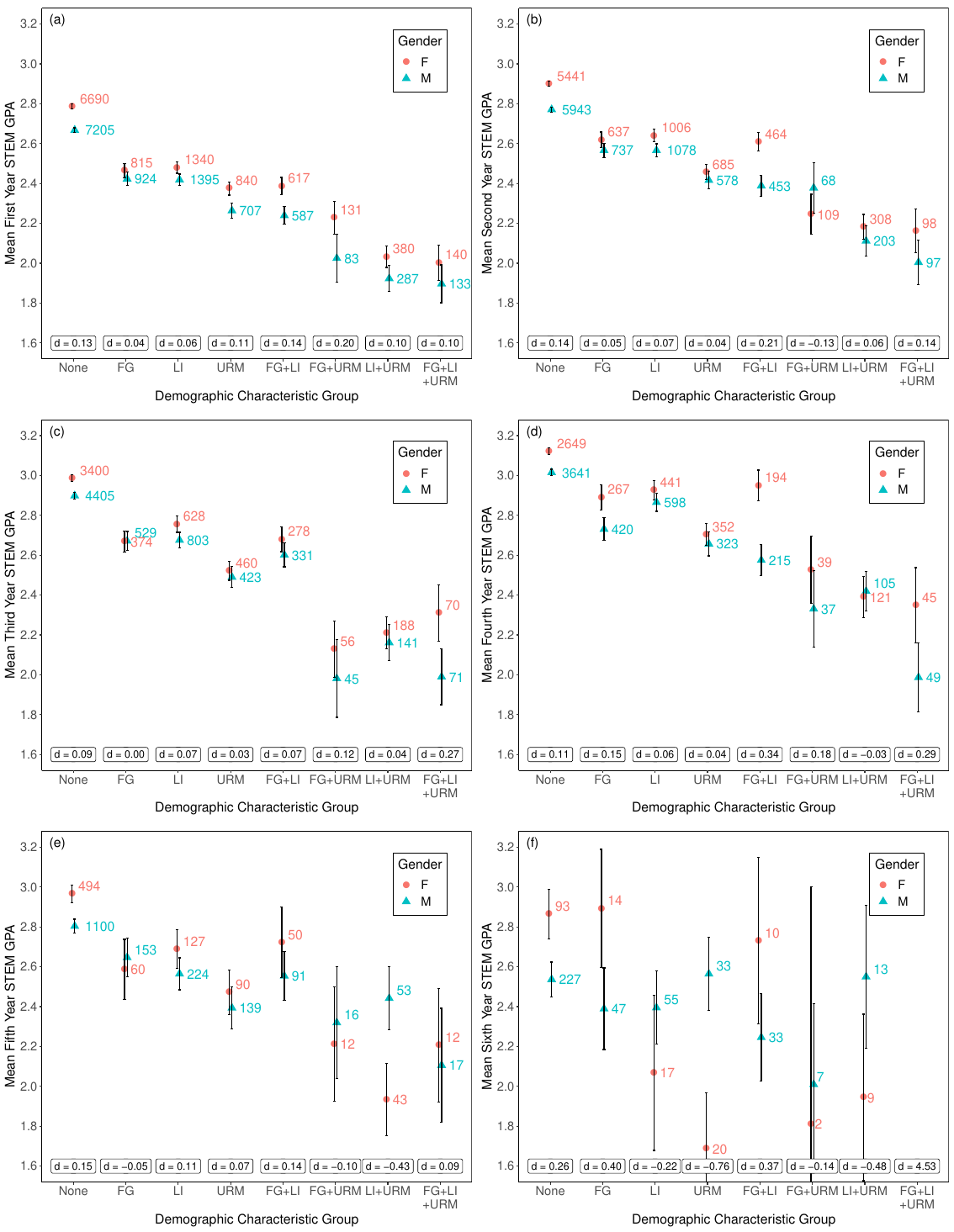}
	
	\caption{\label{figure_gpa_demo_stem}
	Average STEM GPA of each demographic group.
	Students are binned into separate demographic groups based on their status as first-generation (FG), low-income (LI), and/or underrepresented minority (URM) students. The men and women in each demographic group are plotted separately.
	The mean GPA in all courses taken by students in each demographic group is plotted along with the standard error on the mean, with a separate plot for each of the (a) first, (b) second, (c) third, and (d) fourth, (e) fifth, and (f) sixth years.
	The sample size is reported by each point, and Cohen's $d$~\cite{cohen1988} measuring the effect size of the gender difference in each group is reported.}
\end{figure*}

In order to answer \ref{rq_stem}, Figure~\ref{figure_gpa_demo_stem} plots students' mean STEM GPA in a similar manner to Fig.~\ref{figure_gpa_demo_all}.
We note that the general ``dinosaur'' pattern discussed in Fig.~\ref{figure_gpa_demo_all} also holds at least for the first and second years (Figs.~\ref{figure_gpa_demo_stem}a and \ref{figure_gpa_demo_stem}b, respectively).
In the third year and beyond, the general features of the trend continue to hold, with the most privileged students having the highest mean GPA, followed by those with one disadvantage as well as the first-generation and low-income group, followed by the remaining groups of URM students with one or more additional disadvantages.
However, in these later years, the finer details of the plots noted before fall away in favor of a sharper mean GPA decrease for URM students with at least one additional disadvantage in the third year (Fig.~\ref{figure_gpa_demo_stem}c) and a more gradual decrease across all groups in the fourth year (Fig.~\ref{figure_gpa_demo_stem}d) and fifth year (Fig.~\ref{figure_gpa_demo_stem}e).
When restricting the GPA calculations to STEM courses, the sample size becomes too small in the sixth year (Fig.~\ref{figure_gpa_demo_stem}f) to draw meaningful conclusions.

We further observe a trend of students earning higher grades on average in later years, although the rise from the first to the fourth year is somewhat lower in STEM GPA than in overall GPA.
Notably, while in overall GPA this trend seemed to be somewhat universal across demographic groups, in Fig.~\ref{figure_gpa_demo_stem} we see a quicker rise in mean STEM GPA over time for the more privileged students than the less privileged students, particularly comparing the leftmost and rightmost groups.

Regarding gender differences, Fig.~\ref{figure_gpa_demo_stem} shows smaller gender differences in STEM GPA than those observed in overall GPA in Fig.~\ref{figure_gpa_demo_all}.
While in overall GPA women earned roughly 0.2 grade points more than men on average, in STEM GPA that difference is much less consistent and typically ranges from 0 to 0.1 grade points.
For many demographic groups we see no significant differences between men and women's mean STEM GPA.
We do see that there is still a consistent STEM GPA gender difference, albeit smaller than in Fig.~\ref{figure_gpa_demo_all}, among the group of the most privileged students (i.e., those with ``None'' of the disadvantages).
There is also a STEM GPA gender difference among first-generation low-income but non-URM students, however this difference is less consistent and in fact briefly vanishes in the third year.

\subsection{GPA Trends By Major Over Time}

\begin{figure*}
    \centering
    
	\includegraphics[width=0.95\textwidth]{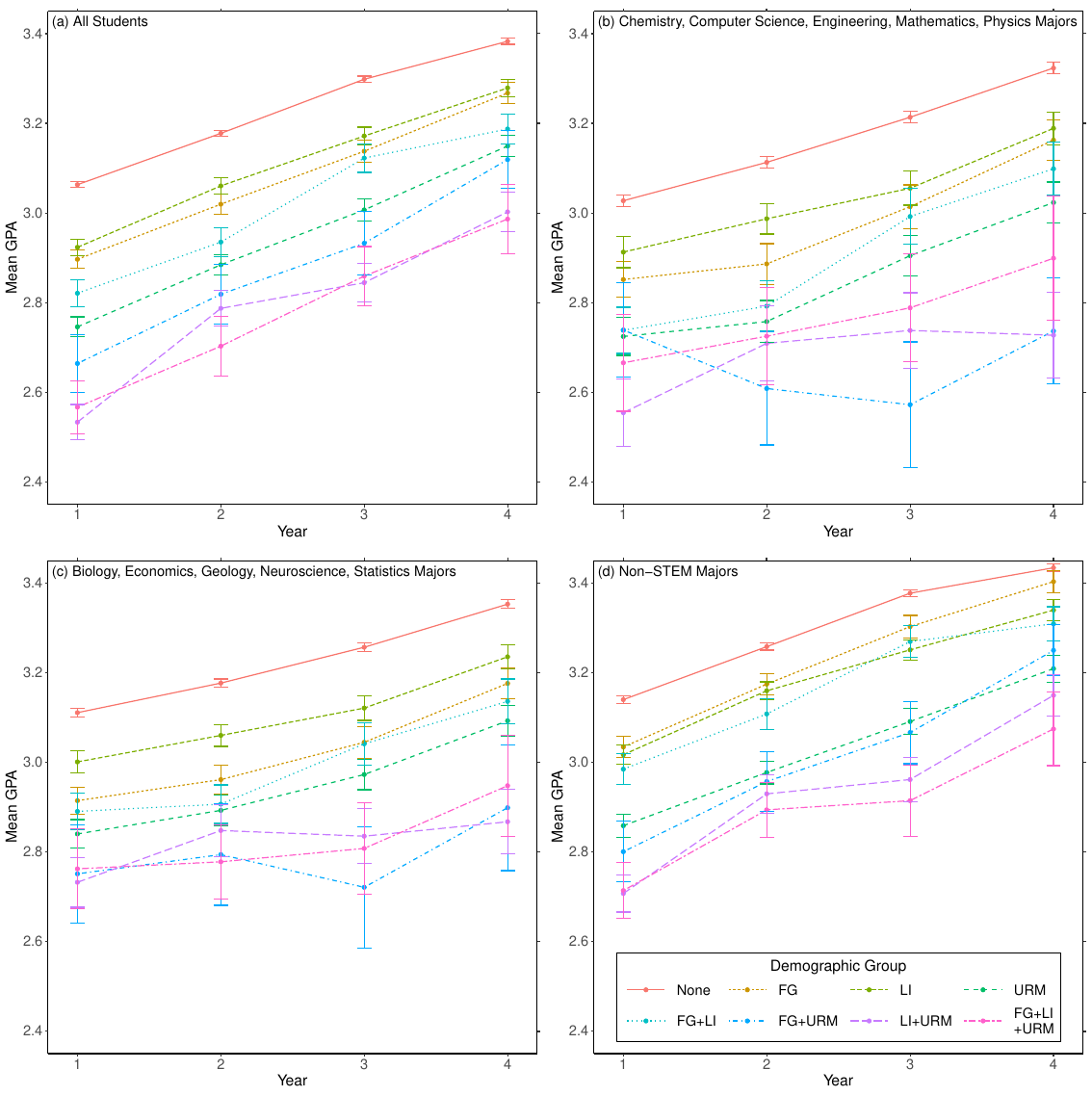}
	
	\caption{\label{figure_gpa_year_all}
	Students are binned into separate demographic groups as in Fig.~\ref{figure_gpa_demo_all}, but not separated by gender. The mean GPA in all courses of each group is plotted over time from year one to four, along with the standard error of the mean. The plots show this for four subpopulations: (a) all students; (b) chemistry, computer science, engineering, mathematics, and physics students; (c) biology, economics, geology, neuroscience, and statistics students; and (d) non-STEM students including psychology.}
\end{figure*}

\begin{figure*}
    \centering
    
	\includegraphics[width=0.95\textwidth]{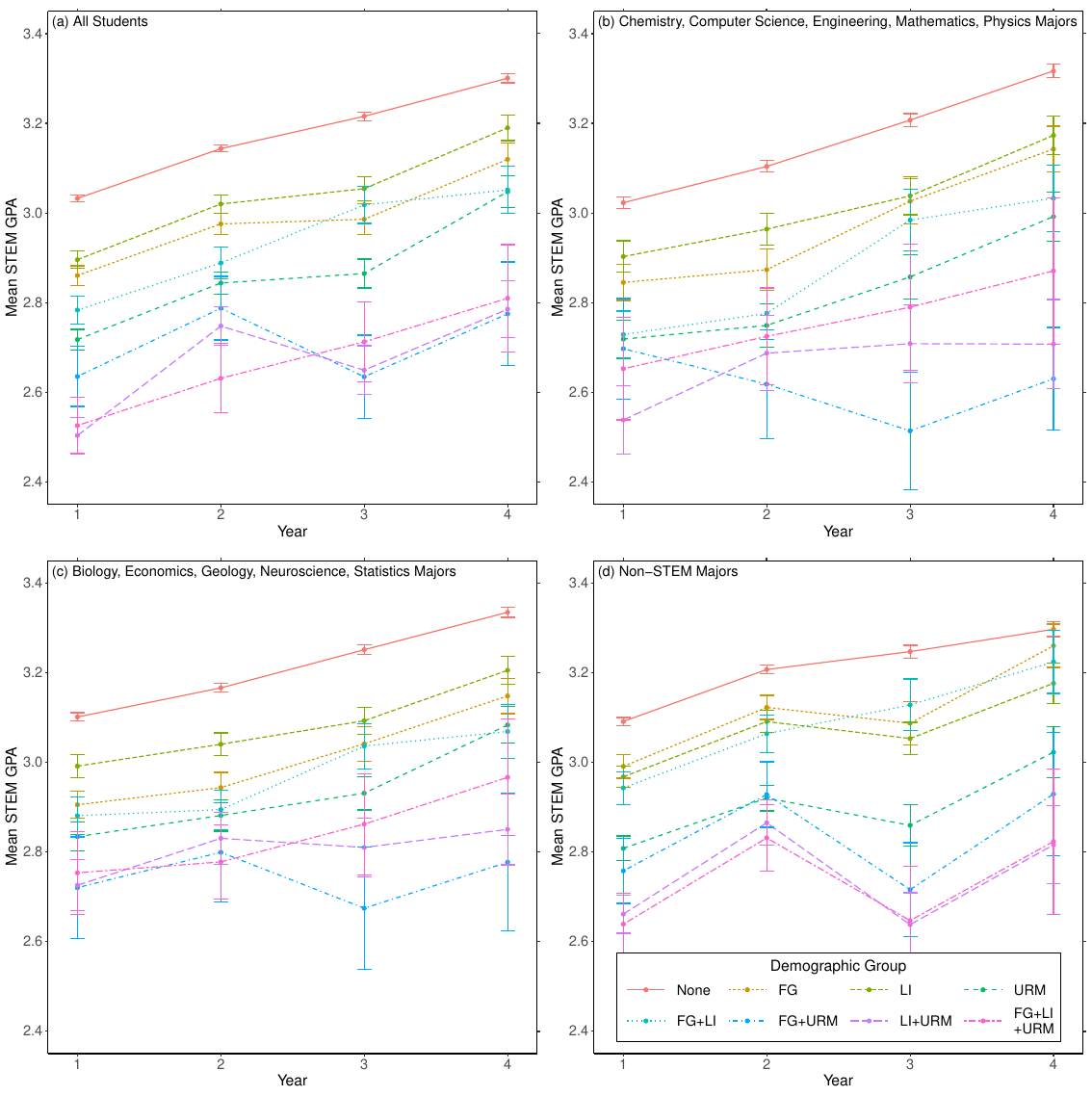}
	
	\caption{\label{figure_gpa_year_stem}
	Students are binned into separate demographic groups as in Fig.~\ref{figure_gpa_demo_stem}, but not separated by gender. The mean GPA in STEM courses of each group is plotted over time from year one to four along with the standard error of the mean. The plots show this for four subpopulations: (a) all students; (b) chemistry, computer science, engineering, mathematics, and physics students; (c) biology economics, geology, neuroscience, and statistics students; and (d) non-STEM students including psychology.}
\end{figure*}

In order to better understand the trends over time in both overall and STEM GPA and answer \ref{rq_time}, we plotted the mean GPA by year in Fig.~\ref{figure_gpa_year_all} and mean STEM GPA by year in Fig.~\ref{figure_gpa_year_stem}.
In these plots, we have not separated men and women and instead focus on the other demographic characteristics while further grouping students into three different groups of majors in order to understand if these trends differ for students in different areas of study.
Further, since the sample size becomes quite small in years five and six for many of the demographic groups of interest, we plot only the mean GPA over the first four years.
In Figs.~\ref{figure_gpa_year_all}a and \ref{figure_gpa_year_stem}a, we plot the mean overall and STEM GPA, respectively, of all students.
In the other subfigures, we plot the mean GPA earned by students majoring in different clusters of majors.
In particular, we plot the mean GPA of engineering (including computer science), mathematics, and physical science (i.e., chemistry and physics) majors in Figs.~\ref{figure_gpa_year_all}b and \ref{figure_gpa_year_stem}b, the remaining STEM majors in Figs.~\ref{figure_gpa_year_all}c and \ref{figure_gpa_year_stem}c, and non-STEM majors in Figs.~\ref{figure_gpa_year_all}d and \ref{figure_gpa_year_stem}d.

These plots make clearer some of the trends noted earlier, especially the rise in mean GPA over time from the first to the fourth year.
However, we can now see that this is not universally true since the first-generation URM students have a drop in mean GPA in the second year for physical science majors (Fig.~\ref{figure_gpa_year_all}b), and in the third year for other STEM majors (Fig.~\ref{figure_gpa_year_all}c).
This trend is even more noticeable in STEM GPA (Fig.~\ref{figure_gpa_year_stem}), where the mean STEM GPA of the group of first-generation URM students drops in the third year for every subpopulation by major.

\section{Discussion}

To start, we consider how much the current system disadvantages students who are first-generation, low-income, or underrepresented minority but not a combination of the two.
Discussing these groups first is helpful in setting the stage for a more complex discussion of the intersectionality of these various demographic characteristics.
We find in Figs.~\ref{figure_gpa_demo_all} and \ref{figure_gpa_demo_stem} that not all of these disadvantages are equal.
In particular, non-URM students who have one disadvantage, namely the first-generation (but not low-income) and low-income (but not first-generation) students, still earn slightly higher grades than even the URM students who are not low-income or first-generation.
Notably, this trend (the ``back'' of the dinosaur plots) is similar in both overall grades (Fig.~\ref{figure_gpa_demo_all}) and in STEM grades alone (Fig.~\ref{figure_gpa_demo_stem}).
The size of this mean grade difference varies from year to year, but in STEM grades it can reach as high as about 0.25 grade points, which at the studied institution is the difference between, for example, a B and B$+$ or B$-$ grade.

The group with the grades most similar to these non-first-generation, non-low-income URM students are the first-generation, low-income non-URM students, who earn both overall (Fig.~\ref{figure_gpa_demo_all}) and STEM (Fig.~\ref{figure_gpa_demo_stem}) grades similar to or very slightly higher than the URM students.
One explanation could be that the lack of resources available due to being first-generation or low-income is not as severe an obstacle as the stereotype threat experienced by URM students.

Turning then to the ``tail'' in the dinosaur plots, we find that consistently the most disadvantaged students in both overall grades (Fig.~\ref{figure_gpa_demo_all}) and STEM grades (Fig.~\ref{figure_gpa_demo_stem}) are the URM students with at least one additional obstacle.
In this case, it appears that the intersection of being low-income and URM is the most disadvantageous combination, with no notable difference in either Fig.~\ref{figure_gpa_demo_all} or Fig.~\ref{figure_gpa_demo_stem} among these students whether or not they are also first-generation.
Meanwhile, the first-generation URM students who are not low-income sometimes have a slightly higher mean GPA than the low-income URM students (Fig.~\ref{figure_gpa_demo_all}).

Another avenue to investigate intersectionality is how gender interacts with the other demographic groups.
Interestingly, in overall GPA (Fig.~\ref{figure_gpa_demo_all}), gender appears to have about the same effect across all demographic groups.
That is, there does not appear to be an intersectional effect of gender identity with other identities as measured by overall GPA.
However, Fig.~\ref{figure_gpa_demo_stem} shows that this is a context-dependent effect, with the gender gap substantially and unevenly reduced across all groups in mean STEM GPA.
For most demographic groups in Fig.~\ref{figure_gpa_demo_stem}, the higher overall GPA earned by women in Fig.~\ref{figure_gpa_demo_all} has vanished completely in STEM GPA.
This is consistent with stereotype threat being the mechanism of disadvantage for women, where stereotypes surrounding STEM disciplines unfairly cause stress and anxiety for women~\cite{astin1993, cross1993, felder1995, felder1998, britner2006, basile2015, cheryan2017, hilts2018}.
Notably, while the gender gap is reduced nearly to zero for most groups in Fig.~\ref{figure_gpa_demo_stem}, there does remain a small consistent gender gap favoring women in the most privileged group of students.
In other groups the gender gap in Fig.~\ref{figure_gpa_demo_stem} is inconsistent across years.
One explanation could be that the wealth of resources available to them may help to alleviate the stereotype threat.

Taking a more temporal view of these GPA trends, Fig.~\ref{figure_gpa_year_all} (overall GPA) and Fig.~\ref{figure_gpa_year_stem} (STEM GPA) have grouped men and women together in order to focus on the other demographic characteristics more closely.
In these plots, the most noteworthy trend is again that, with the sole exception of the first year in Fig.~\ref{figure_gpa_year_all}b, the four groups with the lowest mean GPA (Fig.~\ref{figure_gpa_year_all}) and STEM GPA (Fig.~\ref{figure_gpa_year_stem}) across the first four years are always the four groups containing URM students.
Notably, this trend is true regardless of which group of majors we investigate.
The consistency of this result is particularly striking, showing that the most otherwise disadvantaged non-URM students have fewer obstacles to success than even the most privileged URM students among all students.

Focusing further on the STEM GPA of STEM majors in Figs.~\ref{figure_gpa_year_stem}b and \ref{figure_gpa_year_stem}c, we see that while non-URM students consistently rise in mean GPA over time, the same is not true for all URM students.
In particular, the first-generation URM students who major in chemistry, computer science, engineering, mathematics, or physics  (Fig.~\ref{figure_gpa_year_stem}b) experience a steady decline in mean STEM GPA from year one to two and year two to three.
While the standard error of those means is quite large due to a relatively small sample size, that lack of representation for these students could itself be what is hindering their coursework by causing a stereotype threat.

Based upon the frameworks of critical theory and intersectionality, the main implication of these findings is that many students who come from less privileged backgrounds are not being adequately supported in college in order to catch up with the privileged students~\cite{crenshaw1995, kellner2003, yosso2005, gutierrez2009, taylor2009, tolbert2018, schenkel2020, johnson2012, johnson2017, crenshaw1990, cho2013, mitchell2014, charleston2014, morton2018}. 
The disadvantages of these less privileged students manifest as lower mean overall and STEM GPA for those demographic groups.
In order to promote equity and inclusion, it is crucial that these students are provided appropriate mentoring, guidance, scaffolding, and support in college so that these obstacles can be cleared for students who have been put at a disadvantage relative to their peers through no fault of their own~\cite{birt2019}.
We note that these demographic groups with more disadvantages are likely to consist of students who had K-12 education from schools with fewer resources and less well-prepared teachers than those of the more privileged students, with high school being an especially important time for disadvantages related to STEM learning increasing~\cite{bianchini2003, maltese2011, means2017, bottia2018, daley2019, dou2019}.
Analyses such as those discussed here can help inform the allocation of resources to support these students, with efforts to reduce the classroom stereotype threat of URM students and creating a low-anxiety environment in which all students have a high sense of belonging and can participate fully without fear of being judged being clear priorities.
Additional resources to support low-income and/or first-generation students, e.g., financial support and timely advising pertaining to various academic and co-curricular opportunities, are also important in order to level the playing field and work towards a goal of all students succeeding in college, regardless of their race/ethnicity, socioeconomic status, and parental education history.

\section{Acknowledgments}
This research is supported by the National Science Foundation Grant DUE-1524575 and the Sloan Foundation Grant G-2018-11183.

\bibliography{refs}

\end{document}